# PPTP: Price-based Path-specified Transport Protocol for Named Data Network using Blockchain

Yuhang Ye*, Brian Lee†, Yuansong Qiao*
Software Research Institute, Athlone Institute of Technology
Athlone, Co. Westmeath, Ireland
*{yye, ysqiao}@research.ait.ie; †blee@ait.ie

*Abstract—* Serving as a potential future Internet architecture, Named Data Network (NDN) offers superior information-centric architectural support for mobile ad-hoc networking. Using NDN as an underlying protocol, end-user devices (e.g. IoT device and smart phone) formulate a multi-hop (mesh) network, in which certain devices play a role of forwarding packets for others and/or act as gateways to access the Internet. Nevertheless, an autonomous (selfish) node in an ad-hoc network has two disincentives for forwarding packets for others: energy expenditure and possible delays for its own data. This paper introduces a novel price-based transport protocol –PPTP– for NDN, using blockchain as a payment platform to support money transfers between autonomous nodes thus to incentivise packet forwarding. In PPTP, routers advertise their expected prices for packet forwarding and consumers estimate the costs and select the appropriate paths for content downloading. PPTP is still an on-going project therefore this paper will present the design principle and planed functions, and show how PPTP are related to other existing blockchain-based networking solutions.

*Index Terms—*

## I. INTRODUCTION

The applications of Internet evolved dramatically in the past decades, which are now dominated by content-centric services such as online videos, information retrieval, and social networks. To meet the versatile requirements of modern Internet applications, the host-centric Internet protocol stack is augmented by a number of patches to support mobility, multicast, overlay, and multi-homing. As the result, it becomes clumsy and can hardly offer content-centric services with high-performance [1]. Named-Data Networking (NDN) [2] is a novel solution to rewrite this situation. In NDN, content access is receiver-driven, where consumers only send their content requests (Interest) to get content chunks (Data) back, the network performs name-based forwarding towards any content provider and retrieves the content through the reverse paths to the consumers. Because content is not bound to a host, any authorised node who owns the content can provide it. This design facilitates deploying in-network caches thus improving content delivery efficiencies.

In the conventional Internet, e.g. a consumer downloads content from a content service provider (e.g. Youtube, Netflix and Amazon Prime Video), consumers are billed by Internet Service Providers (ISPs) through monthly contracts, to recover the expenditures of installing/leasing and maintaining network infrastructures, recruit employees and gain revenues. By contrast, when content delivery services are provided by an ad-hoc network, it is critical to figure out how to incentivise autonomous nodes to share their forwarding capability thus to enable content delivery. In literature, monetary incentivisation approaches (e.g. [3]) were proposed for ad-hoc network, aiming at encouraging autonomous nodes to forward packets for others. However, these approaches primarily focus on an adaptive pricing scheme to optimise a global utility function, by assuming consumers and providers are already aware of each other and can capture real-time price information. In practice, this assumption requires an underlying protocol to support. To the best of the authors' knowledge, few existing researches have considered a unified path-with-price discovery solution as the protocol implementation.

In this paper, Price-based Path-specified Transport Protocol (PPTP) is proposed as an implementation to support path-with-price discovery. It inherits our previous transport protocol, namely Path-specified Transport Protocol (PTP). Based on PTP, PPTP further enables routers to advertise the price of using each hop (between two routers) for content forwarding. Then, the price information is piggyback to consumers, together with path information. This allows each consumer to figure out the eligible paths to access the target content and predict the cost of using a certain paths. Because the multipath nature of NDN, a consumer usually can capture more than one paths. To optimise its local utility, each consumer can select certain paths and balance the traffic on them. The payment between consumers and routers are enabled by blockchain and off-chain payment. In particular, a micropayment network is employed to realise frequent off-chain token transfers and blockchain can be used to settle off-chain payments and update account balances.

In recent years, blockchain has been applied to transform versatile Internet services such as data storage (e.g. IPFS [4]) and computation crowdsourcing (e.g. Golem [5]). Also a few teams have started the research on blockchainising network connectivity. For example, New Kind of Network (NKN) [6] introduces a concept of "proof-of-relay" which allows the peers in P2P network to share their forwarding capability and to gain tokens. RightMesh [7] incentivises certain devices that can access the Internet to act as gateways thus allowing other Internet-less devices to surf the Internet. These approaches mainly focus on creating a self-organising network and a charging platform whereas they lack consideration the requirements from a consumer perspective, i.e. a consumer can neither estimate the cost of content downloading, nor decide the preferred paths. PPTP is proposed to fill this gap, to enhance the utility of consumers.

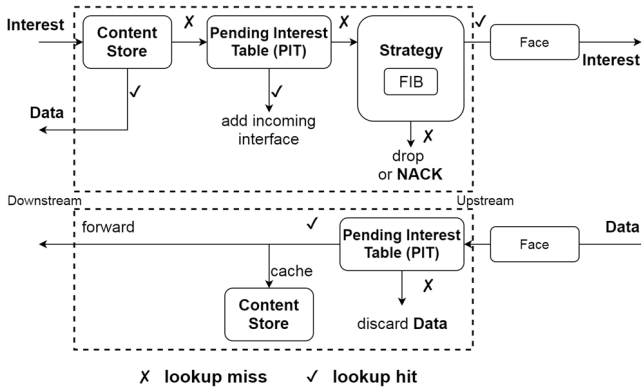

Figure 2 Forwarding process in NDN [2]

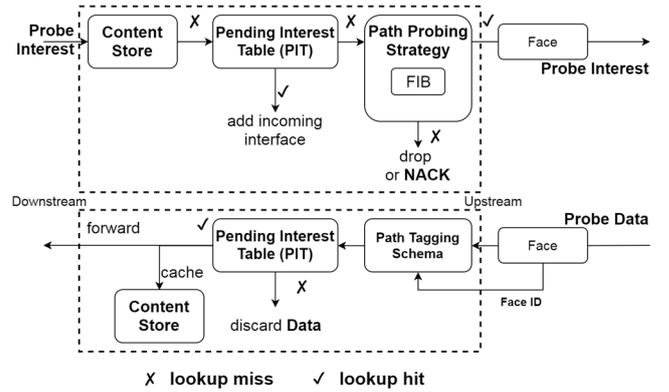

Figure 1 Path probing process in PTP

## II. BACKGROUND

### A. Named Data Network

NDN defines 2 types of packets [2]: Interest and Data. These packets are named by a Uniform Resource Identifier (URI). A client or consumer of content sends an Interest packet in order to fetch content. Once an Interest reaches a data producer, the producer put the corresponding content into the Data packet, and returns it back to the consumer. In this way, the content is not tied to any one host. This addressless routing mechanism enables the direct communication between consumer and any data owner.

### B. Path-Specified Transport Protocol

Our previous work, PTP was developed to enhance the content downloading experiences for NDN consumers. In concept, PTP enables consumers to capture content delivery paths and also to control the traffic on each one. PTP is composed of two stages, namely path probing and path-specified forwarding. During the path probing stage, a new consumer who does not have any prior knowledge about content delivery paths, will discover forwarding information through sending path probing Interest packets. When a probing Interest packet arrives at a producer, an empty tag is created and appended to the probing Data packet that corresponds to the probing Interest packet. This probing Data packet will be returned to the consumer along the Interest's transmission path but in the reverse direction. The tag is appended with the identifier of the receiving interface when the probing Data packet arrives at a router. Finally, the tag is returned to the consumer, and the consumer sets up an independent traffic control module for each tag that corresponds to a path. By appending the tag on some Interest packets, the consumer can direct them to the path thus to control the traffic on it.

### C. PPTP = Price + PTP

NDN is based on a multi-hop request-response model. The request-response feature can be projected to a trading process that a consumer requests and pays for the target content then gets it back. The multi-hop feature can be used to model physical topologies, i.e. each device-to-device connection in physical world corresponds to a link in NDN network. In general, the two features naturally facilitate any end-user who is consuming content delivery services to pay other devices who provide the services.

Nevertheless, a consumer cannot explicit know where it can get the content due to the anycast nature in NDN. In consequence, it brings extreme difficulties to support price-based transmissions as the consumer is not aware of the content downloading path and cannot determine the cost of using it. The introduction of PTP solves this issue by enabling each consumer to discover and control content downloading paths. As a further step over PTP, PPTP allows each consumer collect the price of using each path with its delivery service. Then, blockchain is served as a payment platform to enable token transfer from consumers to providers.

### D. Motivation

The motivation of PPTP is to improve the overall utility of consumers while incentivise other devices to furnish content delivery services. Also PPTP can be used to avoid the network congestion caused by malicious consumers.

#### 1) Utility Improvement and Incentivisation

In an ad-hoc network through spontaneous or impromptu construction, relay nodes are required to forward packets between devices that are not directly connected. However, users are not likely to act as relay nodes because it will degenerates their experiences (e.g. high power consumptions and low application performances). PPTP provides a trading platform to satisfy versatile types of users. For content consumers, PPTP allows them to download content by paying for the service they acquired. For the other nodes that act as relay and formulates transmission paths, PPTP incentivises them to provide routing and forwarding services thus gaining monetary returns from consumers. As a result, it creates a more reliable and reasonable economic equilibrium (i.e. each user optimises its utility function and monetary return) better than an ideal assumption (i.e. each users contribute resource for no reason).

#### 2) Prevent Congestion caused by Malicious Traffic

By billing the usages of network services, PPTP can prevent the congestion caused by malicious requests. In a conventional NDN network, malicious consumers can overly request content and ignore congestion signals thus resulting congestion collapse. In consequence, it affects the QoS of other honest consumers. In PPTP, because every consumer is charged for the network resources that is being consumed, overly requesting content will brings malicious users with monetary costs. If congestion pricing is applied to dynamically adjust the downloading price, it can further give monetary punishments to the malicious nodes who does not react to congestion signals

## III. AN OVERVIEW OF PPTP

PPTP is a transport protocol based on PTP, which extends PTP's path probing and path selecting schemes to discover path prices and to allow consumers to calculate the cost thus selecting the best one(s) for content delivery.

### A. Actors

PPTP has three main actors:

- **Blockchain**: a platform to enable identity registration and verification, payment settlement and dispute resolution.
- **Router**: the nodes that are willing to sell their forwarding capabilities and get monetary returns.
- **Consumer**: the nodes that are willing to get content delivered through a series of routers and pay for it.

### B. Functional Components

PPTP is composed of 5 main functional components:

- **Price probing**: any consumer needs to find the paths that can reach the target content and identify how much does each one cost. Price probing allows consumers to send a special type of requests, i.e. probing request to query the prices of using different links that can be used to access the target content.
- **Price advertising**: price advertising is align with price advertising. When a probing data packet (corresponds to a previous probing request) arrive at a router through a link, the router attaches the price of using this link and its performance measurements to the packet.
- **Path selection**: by evaluating the performance and the price of each path to the target content, a consumer can calculate the utility of each path thus optimising the experience for its application.
- **Off-chain payment**: To use the path to download the target content, consumers send content requests with payment "cheques" to routers. The token inside the cheque will be split and received by the routers who deliver the content that corresponds the content request.
- **On-chain settlement**: through off-chain payment, the tokens gained by each router are kept in a form called commitment transactions. To update the balance of a router, the router will invoke a settlement transaction on the blockchain and submit the final commitment transaction.

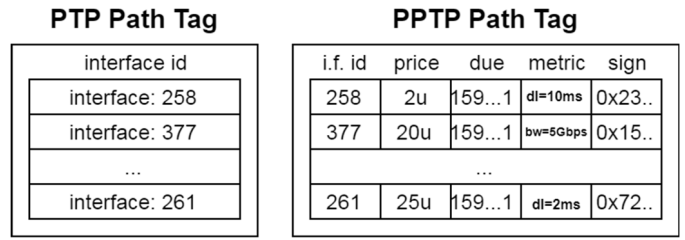

Figure 3 Path Tag in PTP and PPTP

## IV. PATH-WITH-PRICE PROBING

### A. Consumer: Price Probing

To discover the paths to access content and identify how much does each one cost, PPTP requires each consumer to send a probe *Interest* packet to discover a feasible path. When a router receives a probing Interest without a tag, the packet is forwarded to an interface using a round-robin strategy, which distributes packets to interfaces evenly. In the case that the probe Interest is received to a router who does not have a route to access the content, the probe Interest packet is discarded. As a result, it is likely that the consumer can capture all eligible paths to access the content. When a probing Interest packet is forwarded, it does not carry any path or price information. Instead, when the packet is received by a producer which returns a Data packet corresponding to the Interest, the Data packet is attached with a path tag. The tag is the place where path and price information can be attached.

The *path tag* is a stack structure. The different between the path tag in PTP and PPTP is shown in Figure 3. Other than interface identifier, each item in the PPTP tag also contains fields like price, due time, performance metric and a signature of the item. The Data packet carrying a path tag is also known as probing Data, which is returned to the consumer by crossing a series of routers. Any router who received a path tag will append path information to the tag. Finally the path tag will be received by the consumer. This allows the consumer to know the path to downloading the target content.

### B. Router: Price Advertisement

The price advertisement is channel allowing routers to send price information to the consumer. In order to improve pricing flexibility and system scalability, routers do NOT need to submit prices on the blockchain. This prevents significant overheads because routers will need to update price if the market changes. Following the principle of PTP, price advertisement is triggered by the consumer's probing. In particular, when a consumer starts to probe a path and the probe Data packet is crossing any router on the path, the router inserts the price and other information into the path tag.

This loose management of price advertisement results in a chance that a router can confuse others by frequently changing its prices. PPTP avoid this situation by adding a "due" field which regulate routers not to change the price before the due time. If the router advertise multiple versions for the same period, any node in the network can submit these conflict versions to the blockchain. Because each tag item contains a signature, blockchain can easily verify the conflict versions and punish the router who advertise the conflict.

## V. PATH SELECTION AND UTILITY MAXIMISATION

Consumers selects the path that can maximise its return by considering two factors: 1) path price and 2) QoE. This is equivalent to optimising a utility maximisation model at each consumer.

### A. Local Utility Maximisation Model

An abstract formulation of a service discipline and pricing policy allow each user to state the optimality criterion for its required service. The model we proposed here is based on the abstraction method in [8].

Let $s$ denotes a characterisation of the network service received by a consumer and $V_s$ denotes the consumer's satisfaction level, i.e. QoE on service $s$ using a single path. By assuming the charge to the user for the service is $c_s$, the overall utility value returned to the user is given in eq.(1)

$$U_s = \log(V_s) - \log(c_s) = \log\left(\frac{V_s}{c_s}\right) \quad (1)$$

Here, the log function imitates the diminishing marginal utility for both the charge and the QoE return. In addition, maximise $U_s$ is equivalent to maximise $V_s / c_s$ which means to maximise the cost performance ratio when consuming $s$.

For a certain service, the satisfaction level can be modelled based on its QoE requirements. Some typical models of $V(s)$ have been discussed in [8]. For example, voice applications are sensitive to delay thus $V(s)$ can be modelled based on the percentages of packets that are received within 100ms and the inverse of the averaged transmission latency. Following the same principle, it is also feasible to model $V(s)$ for modern video-on-demand applications based on QoS-to-QoE correlation methods [9], i.e. estimating QoE based on bandwidth, latency and packet loss events.

### B. Path Selection

Via path probing, a consumer may capture a number of eligible paths. The consumer can use all of them but need to pay for corresponding usages. For a new consumer participant, it usually cannot know the actual performance of each path. As a result, it will need to predict the utility of each path based on the information provided by routers. This helps the consumer to select certain paths to start downloading. After the consumer starts downloading content from a path, it can measure QoS then estimate the real utility for the path.

Selecting the proper downloading paths from all the paths to improve the overall utility naturally formulates a multi-armed bandit problem. In [10], we presented a heuristic explore-and-exploit method to select the paths with highest bandwidth which can be further extended here to select the paths with highest utilities. In future, we plan to introduce

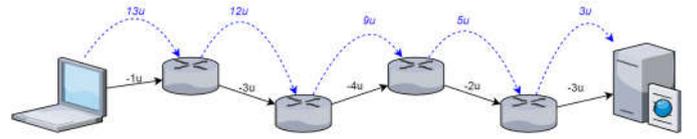

Figure 4 Hop-by-hop Payment

reinforcement learning to solve this multi-armed bandit problem.

## VI. PAYMENT AND SETTLEMENT

The hop-by-hop connectivity amongst routers naturally forms a payment network where consumers pay the routers that deliver content. An example is given in Figure 4, a consumer sent a probe packet and found a 5-hop path to access the content. Via the path tag, the consumer can know the price for each hop (1u, 3u, 4u, 2u, 3u) and it can easily calculate the total price (e.g. 13u) to use this path. To download content from this path, the consumer will pay the first router 13u; the first router will pay the second router 12u, i.e. by taking 1u out of 13u; the second router will pay the third router 9u; … and the last router will pay the content provider 3u. As far as we know, the bottleneck of current blockchain platform is low throughputs (e.g. Ripple XRP can only support around 1500tps), it is not practical to support frequent hop-by-hop online payments. An alternative method is to use off-chain channels, i.e. PPTP will use micropayment networks. Based on the same example as mentioned above, the consumer do not need to set up on-chain transactions in which each one pays the first router 13u. Instead, micropayment enables the consumer and the first router to set up a local token pool, deposit tokens, transfer tokens and confirm their latest account balances using a multi-signature scheme, without the need of blockchain. Later, both of them need to negotiate then select a time point to settle the micropayment using blockchain.

The implementation of off-chain payment is straightforward in NDN. For the payer (e.g. a consumer), it insert a commitment transaction [11] into an Interest packet which will be received by the payee (e.g. a router). If the payee is willing to forward the packet it will accept the commitment transaction. To further forward the Interest packet, it will become a payer and insert another commitment transactions (between it and the next payee) into the Interest packet and forward it to the next payee.

## VII. CONCLUSION

In this paper, we have design a price-based transport protocol –PPTP– for NDN that induce forwarding among autonomous nodes. PPTP allows content consumers to identify the paths that can access target content and calculate the costs of using each one. Blockchain and micropayment network provide a set of channels allows nodes to transfer tokens thus trading content forwarding. Based on the cost and performance metrics of each path, a consumer can estimate utility and adjust the traffic on different paths thus to optimise the overall return. PPTP is a new-born project. In the future, we are going to implement the proposed functional components and design a dynamic pricing scheme thus to improve the network utilisation and enhance the utility for all nodes.